\documentclass[showpacs,preprint,amssymb]{revtex4}
\usepackage{graphicx}% Include figure files
\usepackage{dcolumn}% Align table columns on decimal point
\usepackage{bm}% bold math
\def\be{\begin{equation}} 
\def\ee{\end{equation}}
\def\bea{\begin{eqnarray}} 
\def\eea{\end{eqnarray}}
\def \line{\hbox to \hsize}    
\def\frac #1#2{{#1\over #2}}

\def\psid{\psi^{\dagger}}

\def\ad{a^{\dagger}}

\def\ad{ a^{\dagger}}

\def \ket #1{{\vert #1\rangle}}

\def\1{\mbox{\bf 1}}
\newcommand{\comment}[1]{}

\begin{document}
%\draft %(only for revtex) 

\title{Topological invariants of time reversal invariant superconductors}
\author{Rahul Roy}
\affiliation{University of Illinois, Department of Physics\\ 1110 W. Green St.\\
Urbana, IL 61801 USA\\E-mail: rahulroy@uiuc.edu}

\begin{abstract}
  The topological invariants of gapped time reversal invariant lattice superconductors are studied  by mapping the superconducting mean field Hamiltonian to a Bloch Hamiltonian. There is a single $Z_2 $ invariant in two dimensions and four such invariants in three dimensions. We briefly discuss the properties of states with non-trivial topological invariants. 
\end{abstract}
\maketitle

\section{Introduction}
 
  Various interesting properties of the time reversal symmetry breaking  $p_x + i p_y $ superconductor  arise due to its non-trivial topological properties in momentum space \cite{volovik1989fcs}. A Hopf term in the effective action leads to a quantized value of the Hall conductivity for spin. A topological phase transition is also predicted to occur as the strength of the coupling between fermions in such a system is changed \cite{read2000psf}. This system also supports Majorana fermion excitations which are responsible for the non-abelian statistics of the half quantum vortices  \cite{ivanov2001nas,stone2006fra}. Zero energy edge modes and a chiral current at the edge which lead to a quantized angular momentum of such a system are also expected to occur \cite{stone2004eme}. The non-trivial topology of p-wave states also lies at the heart of the so called ``angular momentum paradox". (For a discussion, see \cite{stone1985pra}). 
  
   In this brief note, we propose to study the topological invariants of ground states of gapped lattice superconductors where time reversal invariance is not spontaneously broken. This work is motivated by recent progress in the topological characterization of band insulators \cite{kane2005qsh,kane2005zto,roy2006zcq,Moore-Balents-topological-invariants-tr-band-structures,roy2006tdt}. We also give examples of superconductors with trivial and non-trivial $Z_2$ indices and state their properties.   In analogy with band insulators, lattice superconductors have also been characterized by Chern numbers \cite{hatsugai2002tqp,hatsugai2004saa}. Such a description is however less useful in the time reversal invariant case.
    
   It was realised from a study of models for the quantum spin Hall effect, that there is a $Z_2 $ invariant in two dimensions which characterizes band insulators, whose connection with the usual Chern numbers leads to robust edge states \cite{kane2005qsh,kane2005zto}. More recently, it was found that there are four such invariants in three dimensions \cite{Moore-Balents-topological-invariants-tr-band-structures,roy2006tdt}. The connection with the Pfaffian matrix and surface states was discussed in \cite{fu2006tit}. Here, we show how one can adapt these results to characterize gapped time reversal invariant lattice superconductors \footnote{This has also been independently suggested \cite{skroy}}.

\section{Invariants for Superconductors}  
   Superconductivity arises from Cooper pairing of fermions  and can occur in the presence of an arbitrarily weak attractive interaction between them. It is usually accompanied by the formation of a gap at the Fermi surface, though this gap can vanish at specific points or lines of points. Here we shall consider multi-band superconductors which are known to have interesting properties such as novel collective modes \cite{leggett1966npf} and ``interior gap superfluidity" \cite{liu2003igs}.
      
    To capture the pairing between the particles, one can write down a mean field BdG Hamiltonian of the form 
  \bea
   H_{Bogoliubov} =  H_{ij} \ad_i a_j + (\Delta_{ij}\ad  _i \ad _j + \Delta_{ij}^{*}a_j a_i ) 
  \eea 
  The form of the gap can be determined self consistently from the microscopic Hamiltonian from which the mean field BdG Hamiltonian is derived. 
  Here we shall assume that the gap and hence the BdG Hamiltonian can take various functional forms in momentum space, and study the consequences of its global topology without worrying about how such a mean field Hamiltonian might arise from a microscopic Hamiltonian. 
 
  In a periodic potential, a multi-band BdG Hamiltonian can be written as 
  \bea
   H = \int  dk \,\, \left[h_{\alpha,\beta}(k) \ad_{\alpha,k} a_{\beta,k} + (\Delta_{\alpha,\beta}(k)\ad  _{\alpha,k} \ad_{\beta,-k}  + h.c ) \right]
  \eea
  where $k = (k_1,..,k_n) ,\, dk = \prod_{i=1} ^n dk_i $\,, and $ a_{\alpha, k}, \ad_{\alpha,k} $ are annihilation and creation operators corresponding to Bloch states with band indices $\alpha, \beta $.  
 When time reversal invariance is a symmetry of the system, we can write the Hamiltonian in terms of  sets of time reversed pairs of bands $ \ket{\gamma(k)},\ket{\omega(k)} $ such that the operation of time reversal on eigenstates which belong to these  bands is $ \Theta \lambda (\ket{\gamma(k)}) = \lambda^* \ket{\omega(-k)}, \Theta \lambda(\ket{\omega(k)}) = - \lambda^* \ket{\gamma(-k)}$ for an arbitrary complex number $\lambda$. 
  
   Using the Nambu formalism, we can write the Hamiltonian as 
   \bea 
    H_{BdG} = \int \,dk\, \left[ \psid(k)\, \hat{H} \,\psi(k) + {1 \over 2} \textrm{tr}(H) \right]
    \eea
    where $ \psi^{T}(k) = (a_{\alpha,k}\,\, \ad_{\alpha,-k} )  $ and 
  \begin{eqnarray*}
  \hat{ H} & = & \left(
   \matrix{ h_{\alpha,\beta}(k) & \Delta_{\alpha,\beta}(k) \cr  {\Delta}^{\dagger}_{\alpha,\beta} (k) & -h^{T} _{\alpha,\beta}(-k) } \right)   \\
   \end{eqnarray*} 
  
   With the above BdG Hamiltonian, one can associate a Bloch Hamiltonian of the form
 \bea
 H_{Bl} = \int \,dk\, \phi ^{\dagger}(k)\, \hat{H} \,\phi(k)
 \eea
  where $ \phi^T(k) = (a_{\alpha,k}\,\, \tilde{a}_{\alpha,k} ) $ where $ \tilde{a}_{\alpha,k}$ are annihilation operators corresponding to  a new set of Bloch bands equal in number to the original set of $\{\alpha\}$.
  
  We consider a system which has a non-vanishing gap. The superconducting ground state then is the one that corresponds to all the negative energy eigenstates of the Bloch Hamiltonian $H_{Bl}$ being filled.  
  
   It was shown in \cite{Moore-Balents-topological-invariants-tr-band-structures,roy2006zcq,roy2006tdt} that a set of $Z_2 $ indices can be defined for such a system, which stay invariant when the Hamiltonian is adiabatically changed in such a way that the gap never vanishes. 
   The extra structure of the BdG Hamiltonian, namely the skew symmetry of $\Delta$ and the form of $H_{BdG}$ does not change this result. 
   Thus we see that in two dimensions, there is a single $Z_2 $ invariant which can be used to characterize gapped time reversal invariant lattice superconductors, while in three dimensions, there are four such invariants.  
  
   In two dimensions, the $Z_2 $ invariant can be calculated by the formula : 
   \bea
  E = |\sum_{c_n > 0 } c_n | \textrm{mod }2
  \eea
  where $ c_n $ are the Chern numbers of the filled bands of $H_{Bl}$ and the sum is taken over the set of bands which have positive Chern numbers. In three dimensions, there are four $Z_2$ invariants. For a cubic Brillouin zone, one each can be associated with three distinct faces, while the fourth one characterizes the trapped monopole charge in momentum space.  
   Ordinary s-wave superconductors in two and three dimensions belong to the class with all $Z_2$ indices zero. 
   
    A simple   2-d tight binding model with a non-trivial $Z_2 $ index for  the  p-wave state can be written as  
  \bea
   h_{\alpha,\beta}(k) = [-2t(\cos(k_x b) + \cos(k_y b) )- \mu]\delta_{\alpha,\beta} \\
   \Delta(k) = 2 i \Delta_x (\sin(k_x b)I  + i \sin(k_y b)  \sigma_3 ) 
  \eea   
  where the matrix indices correspond to spin components along the z axis.
    The low energy effective Hamiltonian shows that this corresponds to two p-wave superfluids with order parameters $ p_x + i p_y $ and $p_x - i p_y $ for the two spin components. Since the $Z_2 $ index is intimately connected with the Chern numbers, our analysis shows that the above state also has the non-trivial properties of a $p_x + i p_y $ superfluid. Thus this state will also have half quantum vortices with non-abelian statistics as well as robust edge states though the net current and the angular momentum will be zero.
    In contrast, the superconducting $ d_{x^2-y^2} + i d_{xy}$ state \cite{laughlin1998mid}  with gap function   
        \begin{equation}
    \Delta(k) = \Delta_{x^2-y^2}[\cos(k_x b) - \cos(k_y b)] I + \\
     i\sigma_{3} \Delta_{xy}\sin(k_x b)\sin(k_y b)
    \ee  
   is topologically equivalent to the s-wave state with no edge states. 
   
   In three dimensions, an exotic superconductor analogous to the exotic insulator with trapped monopoles in momentum space \cite{roy2006tdt} is also possible. The three dimensional superconductors with non-trivial $Z_2$ indices will have various surface states analogous to the ones in \cite{fu2006tit}. 
Models for the exotic superconductor can be also be written down analagous to the insulator states.

   I would like to express my gratitude to Prof. Michael Stone and Prof. Sheldon Katz for extensive discussions which have greatly helped my understanding of the subject and the Deparment of physics at the University of Illinois for support.
\bibliography{qshe}
\end{document}